\documentclass[pla,preprint,onecolumn]{elsarticle}

\usepackage{hyperref}
\hypersetup{colorlinks=true, urlcolor=red, citecolor=black, linkcolor=black}
\usepackage{amsmath,amssymb,braket,subfloat,physics}
\usepackage{graphicx,epstopdf}
\usepackage{subfigure}
\usepackage{epstopdf}
\usepackage{lastpage}
\begin{document}
\title{A comparison between higher-order nonclassicalities of SUP-engineered coherent and thermal states}
\author{Deepak}
\ead{deepak20dalal19@gmail.com}
\author{Arpita Chatterjee\corref{cor1}}
\ead{arpita.sps@gmail.com}
\cortext[cor1]{Corresponding author}
\date{\today}
\address{Department of Mathematics, J. C. Bose University of Science and Technology,\\ YMCA, Faridabad 121006, India}

\begin{abstract}

We consider an experimentally obtainable SUP operator, defined by using a generalized superposition of products of field annihilation ($a$) and creation ($a^\dagger$) operators of the type, $A = saa^\dagger+t{a^\dagger}a$ with $s^2+t^2=1$. We apply this SUP operator on coherent and thermal quantum states, the states thus produced are referred as SUP-operated coherent state (SOCS) and SUP-operated thermal state (SOTS), respectively. In the present work, we report a comparative study between the higher-order nonclassical properties of SOCS and SOTS. The comparison is performed by using a set of nonclassicality witnesses (e.g., higher-order antiubunching, higher-order sub-Poissonian photon statistics, higher-order squeezing, Agarwal-Tara parameter, Klyshko's condition). The existence of higher-order nonclassicalities in SOCS and SOTS have been investigated for the first time. {\color{black} In view of possible experimental verification of the proposed scheme, we present exact calculations to reveal the effect of non-unit quantum efficiency of quantum detector on higher-order nonclassicalities.}

\end{abstract}
\maketitle
\section{Introduction}
\label{sec1}

It is a well-known fact that the superposition principle lies at the heart of quantum mechanics \cite{vladimir}. Being a linear theory, quantum mechanics allows the existence of various types of superposition states. However, linearity is not a unique property of quantum mechanics only. Even in classical optics, interference is being observed which is a clear evidence of superposition principle via first order correlation \cite{aram}. But first order correlation or first order interference are not strong enough to manifest some unique (nonclassical) properties of quantum mechanics which have no classical analogue. To realize the actual beauty, power and mystery of quantum mechanics, we need to study the second or higher-order correlations.

A quantum state is identified as nonclassical if the corresponding Glauber-Sudarshan $P$-function has negative values \cite{roy,ecg}. But unfortunately, except a single proposal for the measurement of the $P$-function in a special case, there is no experimental method for determining the $P$-function \cite{kiesel}. Consequently, several operational measures to witness the nonclassicality have been developed (\cite{adam} and references therein). Any finite set of these witnesses of nonclassicality is only sufficient \cite{rich}. So, failure of any of these criteria does not imply classicality of the state, but satisfying a condition signifies nonclassicality. Almost all of these witnesses can be expressed as the moments of annihilation and creation operators, and a nonclassical property detected by a second-order correlation (i.e., a moment-based criterion involving terms up to fourth order in annihilation and creation operators) is called a lower-order nonclassical property. Naturally, higher-order nonclassicality refers to the nonclassical features related to higher-order correlations. Most frequently used higher-order nonclassical features are higher-order antibunching (HOA) \cite{pathak1}, higher-order sub-Poissonian photon statistics (HOSPS) \cite{hari,amit}, higher-order squeezing (HOS) of Hillery type \cite{mark} and Hong-Mandel type \cite{hong}. All of these nonclassical features have lower-order counterparts which have been studied more rigorously in \cite{pathak2,dodonov}. In addition to these, Agarwal-Tara \cite{girish, priya} and Klyshko's \cite{klyshko} conditions are two measures of utmost interest. Recently, many investigations have been reported in the direction of higher-order nonclassicalities because of a large number of successful experimental characterization of such states \cite{aven,allevi1}. Also the fact that higher-order criteria can identify weak nonclassicalities, not detected by their lower-order counterparts, has turned this area a promising field of research \cite{allevi2,jan}.

The generation and manipulation of nonclassical states of the electromagnetic field, at the single-photon level, have earned the attention of quantum optics
and quantum information community \cite{kok,braun}, in the context of physical realization of quantum tasks, protocols and communications \cite{zeilinger,bouwn,bouwn1,nielsen} using continuous quantum variables. Different methods for preparing nonclassical states have been suggested, based on operations of photon addition ($a^\dagger$) \cite{zavatta} and subtraction ($a$) \cite{wenger} on a classical field. Agarwal and Tara \cite{agar} first proposed an $m$-photon-added scheme to create a nonclassical state from any classical state. Zavatta
et. al. \cite{zavatta1} demonstrated a single photon-added coherent state by homodyne tomography technology. In another real scheme, Zavatta et. al. \cite{zavatta2} used single-photon interference for implementing superpositions of distinct quantum operations. Hu et. al. \cite{hu} investigated
the nonclassical properties of the field states generated by subtracting any number of photons from a squeezed thermal state. Lee and Nha \cite{lee} studied the action of an elementary coherent superposition of $a$ and $a^\dagger$ on continuous variable systems. A SUP operator $saa^\dagger+t{a^\dagger}a$ is proposed in \cite{ahr}, where $s$ and $t$ are scalars with $s^2+t^2=1$. This experimentally feasible operation introduced nonclassicality in the classical coherent state and non-Gaussianity in the classical thermal state. The nonclassicality in the SUP-operated coherent (SOCS) and thermal (SOTS) states can be quantified and analysed using quasiprobability distributions in the phase space. However, to the best of our knowledge, no effort has been made so far to investigate the higher-order nonclassical properties of SOCS and SOTS. In this paper, we concentrate on higher-order nonclassicalities of SUP-operated coherent and thermal states.

Apart from what stated above, the work is also motivated by the success of detecting higher-order nonclassicality in laboratory using various technologies \cite{allevi1,allevi2}. Recently, HOA, HOSPS and HOA have been reported in optomechanical and optomechanical-like systems \cite{nasir1}, finite dimensional coherent state \cite{nasir2}, most general form of the hyper Raman process \cite{kishore2}, photon added and subtracted squeezed coherent states \cite{kishore3} etc. In addition, applications of nonclassical states \cite{agar1} and effect of non-Gaussianity inducing operations (e.g., photon addition, photon subtraction, photon addition followed by subtraction and vice-versa) on the nonclassical properties \cite{kishore3} have been studied in detail in the recent past. Keeping the above in mind, we investigate the possibilities of observing HOA, HOSPS, HOS of Hong-Mandel type etc. in the SUP-operated coherent and thermal states. {\color{black} We also discuss how imperfection in single-photon detector \cite{gisin} influences the characterization of higher-order nonclassicalities of SOCS and SOTS.}

The structure of the paper is as follows: in Sec.~\ref{sec2}, we have introduced SOCS and SOTS and corresponding normalization constants. In Sec.~\ref{sec3}, we have checked the
existence of various higher-order nonclassicalities in the superposition states described in the previous
section. Finally, the article is concluded in Sec.~\ref{sec4}.

\section{Quantum states under consideration}
\label{sec2}

%
%

In this section, we focus on SUP-operated coherent and thermal states. Being an eigenstate of the annihilation operator $a$, the coherent state corresponding to the unique eigenvalue $\alpha$, can be simply written as $a\ket{\alpha}=\alpha\ket{\alpha}$. On  the other hand, the Fock state representation of a single-mode thermal field at an absolute temperature $T$ and with frequency $\omega$ is $\rho_{n}=\sum_n a_n\ket{n}\bra{n}=\frac{1}{1+\bar{n}}\sum_n^{}\Big(\frac{\bar{n}}{1+\bar{n}}\Big)^n\ket{n}\bra{n}$, where $\bar{n} =[e^{h\omega/kT}-1]^{-1}$ is the mean photon number, $k$ being the Boltzmann constant. Given a coherent (thermal) state as an input field, the SUP-operated coherent (thermal) state can be described as \cite{lee}

\begin{eqnarray}
\left.
\begin{array}{lcl}
\label{eq1}
\ket{\psi_1} & = & N_1^{-1/2}A\ket{\alpha},\\\\
\rho_{\psi_2} & = & N_2^{-1}A^\dagger\rho_nA,
\end{array}
\right\}
\end{eqnarray}
where $A=saa^\dagger+t{a^\dagger}a$, $s^2+t^2=1$ is the SUP operator, $N_1 = \Big[s^2 +(1+2s^2+4st)|\alpha|^2 +(1+2st)|\alpha|^4\Big]$ and $N_2 = \Big[s^2(1+\bar{n})(1+2\bar{n})+4st\bar{n}(1+\bar{n})+t^2\bar{n}(1+2\bar{n})\Big]$ are the normalization constants for SOCS and SOTS, respectively \cite{ahr}.\\

\section{Nonclassical features of the superposed states}
\label{sec3}

Any quantum state $\ket{\psi}$ is referred as nonclassical if the corresponding Glauber-Sudarshan $P$-function is not a classical probability distribution \cite{roy,ecg}.
That means the state having the negative $P$-function is not of classical nature and can be considered as a nonclassical one. As $P$ function is very rarely detectable in laboratories \cite{kiesel}, a number of experimentally measurable nonclassicality witnesses like Mandel's $Q_M$ parameter, higher-order antibunching (HOA), higher-order sub-Poissonian photon statistics (HOSPS), higher-order squeezing (HOS) of Hong-Mandel type, Agarwal-Tara and Klyshko's conditions have been developed. These operational
criterion for identifying nonclassicality can be expressed as the moments of annihilation and creation operators \cite{nasir2,kishore3}. To utilize these witnesses for detecting the nonclassicalities of SOCS and SOTS, it is better to compute an analytic expression for the most general moment $\langle{a^{\dagger m} a^n}\rangle$, where $m$ and $n$ are the non-negative integers. This is the most general moment in the sense that any other moment can be derived as a special case of it. {\color{black}Using the algebraic properties of the boson operators \cite{whl}
\begin{eqnarray*}
[a, a^{\dagger l}] = \frac{\partial a^{\dagger l}}{\partial a^\dagger}
\end{eqnarray*}
and
\begin{eqnarray*}
[a^\dagger, a^l] = -\frac{\partial a^{l}}{\partial a}
\end{eqnarray*}
where $l$ is an integer, we obtain
\begin{eqnarray}\nonumber
\label{eq4}
a a^{\dagger p}a^q a^\dagger & = & aa^{\dagger p}(q a^{q-1}+a^\dagger a^q)\\\nonumber
& = & q aa^{\dagger p}a^{q-1}+aa^{\dagger {p+1}}a^q\\
& = & q(p a^{\dagger{p-1}}+a^{\dagger p} a)a^{q-1}+\big\{(p+1) a^{\dagger p}+a^{\dagger{p+1}} \big\}a^q\\\nonumber
& = &  pq a^{\dagger {p-1}}a^{q-1} + (p+q+1)a^{\dagger p}a^q + a^{\dagger {p+1}}a^{q+1}
\end{eqnarray}}

Thus the expectation of $a^{\dagger m}a^n$ with respect to SOCS is given by
\begin{eqnarray}\nonumber
\label{eq4}
{\langle{a^{\dagger m}a^n}\rangle}_1 & = & N_1^{-1}\bra{\alpha}A a^{\dagger m}a^n A\ket{\alpha}\\
& = & N_1^{-1}\alpha^{* m}\alpha^n\big[(s+t)^2\{mn+(m+n+1)|\alpha|^2+|\alpha|^4\}\\\nonumber
& & +s(s+t)(m+n+2|\alpha|^2)+s^2\big],
\end{eqnarray}
while with respect to SOTS, it is
\begin{eqnarray}\nonumber
\label{eq7}
{\langle{a^{\dagger m}a^n}\rangle}_2 & = & N_2^{-1}\,\mathrm{tr}(a^{\dagger m}a^nA\rho_r A)\\
& = & \delta_{mn}N_2^{-1}\sum_r \frac{a_r~r!}{(r-n)!}(s+(s+t)r)^2,
\end{eqnarray}
where $\delta_{mn}$ is the usual delta function.

\subsection{Higher-order photon statistics}

The Mandel's parameter $Q_M$ \cite{mandel} illustrates the nonclassicality of a quantum state through its photon number distribution. The introductory definition of $Q_M$ can be generalized to an arbitrary order $l$ as \cite{sanjib}
\begin{eqnarray}
\label{eq12}
Q_M^{(l)} & = & \frac{\langle{(\Delta{\mathcal{N}})^l}\rangle}{\langle{a^\dagger a}\rangle}-1,
\end{eqnarray}
where $\Delta{\mathcal{N}}\,=\,a^\dagger a-\langle{a^\dagger a}\rangle$ is the dispersion in the number operator $\mathcal{N}=a^\dagger a$. Using the identity \cite{sanjib}
\begin{eqnarray*}
\langle{(\Delta{\mathcal{N}})^l}\rangle = \sum_{k=0}^l {l \choose k}(-1)^k\langle (a^\dagger a)^{l-k}\rangle{\langle a^\dagger a\rangle}^k
\end{eqnarray*}
and \cite{moya1}
\begin{equation}\nonumber
(a^\dagger a)^r = \sum_{n = 0}^r  S_r^{(n)}a^{\dagger n}a^n,
\end{equation}
where $S_r^{(n)}$ is the Stirling number of second kind \cite{stegun}
\begin{equation}\nonumber
S_r^{(n)} = \frac{1}{n!}\sum_{j=0}^n (-1)^{n-j}{n\choose j}j^r,
\end{equation}
the higher-order Mandel's parameter $Q_M^{(l)}$ can be evaluated explicitly up to order $l$. The negativity of $Q_M^{(2)}$ signifies the negativity of the conventional Mandel's $Q_M$. All expectations in \eqref{eq12} have been calculated with help of \eqref{eq4} and \eqref{eq7}.

\begin{figure*}[htb]
\centering
\includegraphics[scale=1]{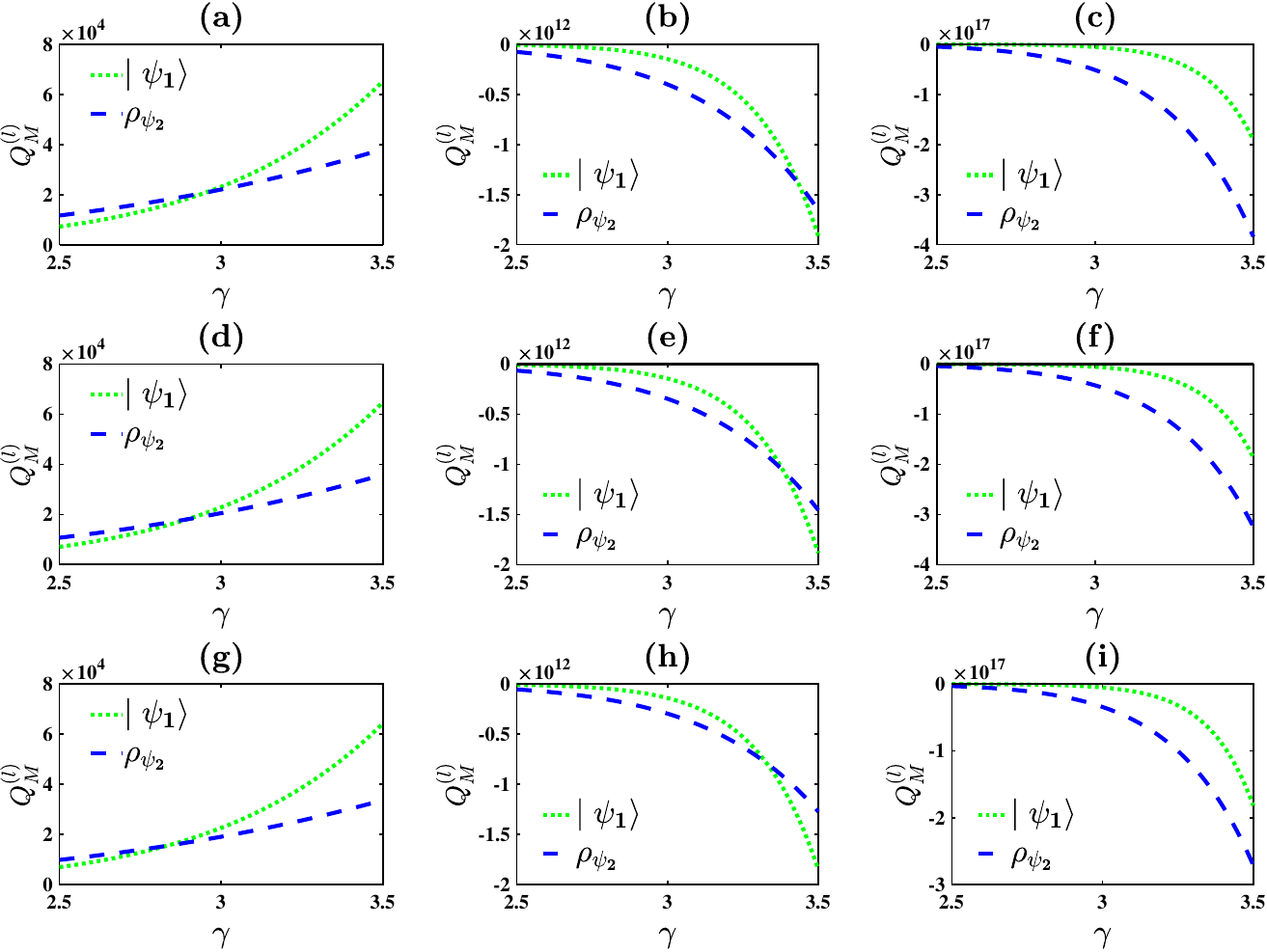}

\caption{(Color online) Comparison of $Q_M^{(l)}$ with respect to the variable $\gamma$ which is equal to  $|\alpha|$ for coherent state $\ket{\psi_1}$ and $\bar{n}$ for thermal state $\rho_{\psi_2}$ and with (a)-(c) $s=0.2$ and $l=2,5,7$, (d)-(f) $s=0.5$ and $l=2,5,7$, (g)-(i) $s=0.8$ and $l=2,5,7$, respectively.}
\label{fig1}
\end{figure*}
\begin{figure}[htb]
\centering
\includegraphics[scale=1]{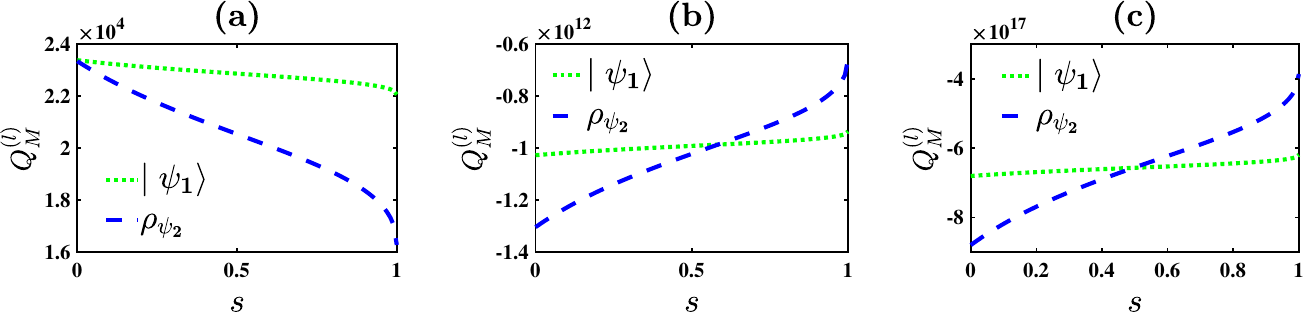}
\caption{(Color online) Comparison of $Q_M^{(l)}$ as a function of $s$ and with (a) $|\alpha|=\bar{n}=3$ and $l=2$, (b) $|\alpha|=\bar{n}=3.37$ and $l=5$,  (c) $|\alpha|=\bar{n}=3.69$ and $l=7$, respectively.}
\label{fig2}
\end{figure}

The negative values of $Q_M^{(l)}$ parameter is a sufficient condition \cite{priya} for the negativity of the $P$ function and hence it gives a witness for nonclassicality. The photon number distribution is classified as sub-poissionian, poissionian or superpoissionian if the Mandel's parameter $Q_M^{(l)}$ for any $l>1$, is less than, equal to or greater than zero, respectively. In Fig.~\ref{fig1}, a comparison between Mandel's criteria of a SUP-operated coherent state $\ket{\psi_1}$ with respect to the state parameter $\alpha$ and a SUP-operated thermal state $\rho_{\psi_2}$ with respect to the state parameter $\bar{n}$ is shown for different values of $s$. It is clear that the nonclassicality, not detected by the lower-order parameter $Q_M^{(2)}$, can be identified by the higher-order criteria (when $l=5$ or $7$). {\color{black}Thus the presence of higher-order nonclassicality while its lower-order counterpart is absent proves the
relevance of the present study}. It is also notified that SOTS is more nonclassical than SOCS in a certain range of $\alpha$ or $\bar{n}$ and after crossing it, SOCS displays more nonclassicality than SOTS. The upper limit of this certain range increases from $\alpha$ or $\bar{n}\approx 3.5$ to 3.8 as $l$ increases from $5$ to $7$, keeping $s$ fixed. In Fig.~\ref{fig2}, $Q_M^{(l)}$ is plotted as a function of $s$ and for different values of $\alpha$, $\bar{n}$ and $l$. Here SOCS shows mere changes in nonclassicality with respect to $s$ while a notable change is observed in case of SUP-operated thermal state.

\subsection{Higher-order antibunching}

HOA is another criteria which can be used to check the nonclassicality of any quantum state. Lee introduced HOA by using the concept of majorization \cite{ching}. Later Pathak and Gracia \cite{pathak1} modified this to give a clear physical meaning and to simplify the expression.
The $(l-1)$-th order antibunching of a quantum state is given as following
\begin{equation}
\label{eq13}
{\color{black}d_h^{(l-1)}}=\langle a^{\dagger l}a^l\rangle -{\langle a^\dagger a\rangle}^l\,\, <\,\,0
\end{equation}
Since the negativity of $d_h^{(l-1)}$ indicates that the probability of photons coming bunched is less compared to that of coming independently, therefore the nonclassicality feature \eqref{eq13} typifies how suitable a state $\ket\psi$ is as a single photon resource. In this subsection, we have calculated the values of $d_{h}^{(l-1)}$ for SOCS and SOTC using \eqref{eq4} and \eqref{eq7}
and plotted them. The signature of lower-order antibunching can be obtained as a special case of \eqref{eq13} for $l = 2$ and for $l\ge 3$, the negative values of $d_{h}^{(l-1)}$ correspond to the higher-order antibunching of order $(l-1)$.

\begin{figure*}[h]
\centering
\includegraphics[scale=1]{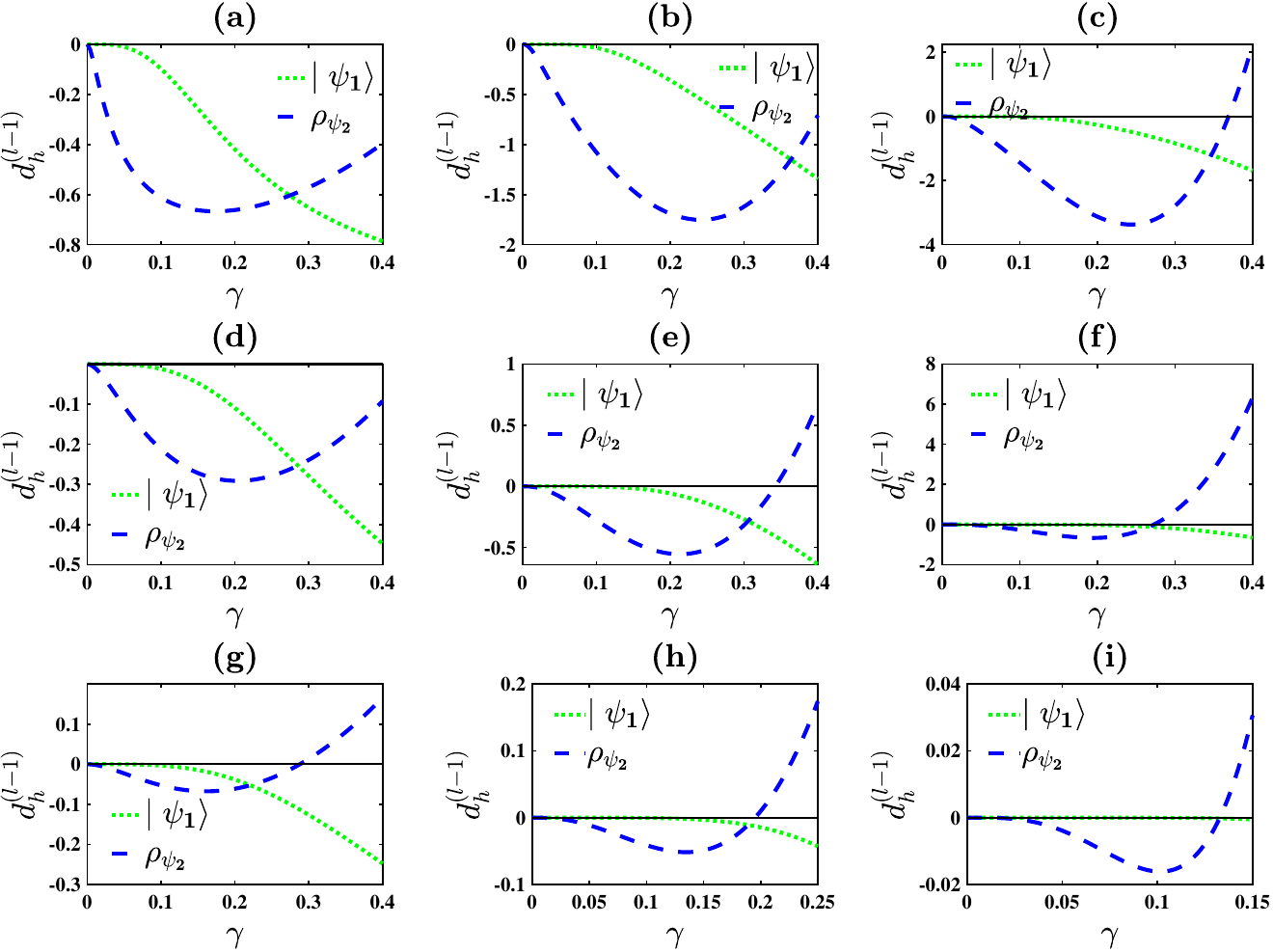}
\caption{(Color online) Comparison of $d_h^{(l-1)}$ with respect to the variable $\gamma$ which is equal to $|\alpha|$ for coherent state $\ket{\psi_1}$ and $\bar{n}$ for thermal state $\rho_{\psi_2}$ and with (a)-(c) $s=0.2$ and $l=2,3,4$, (d)-(f) $s=0.5$ and $l=2,3,4$, (g)-(i) $s=0.8$ and $l=2, 3, 4$, respectively.}
\label{fig3}
\end{figure*}
\begin{figure}[h]
\centering
\includegraphics[scale=1]{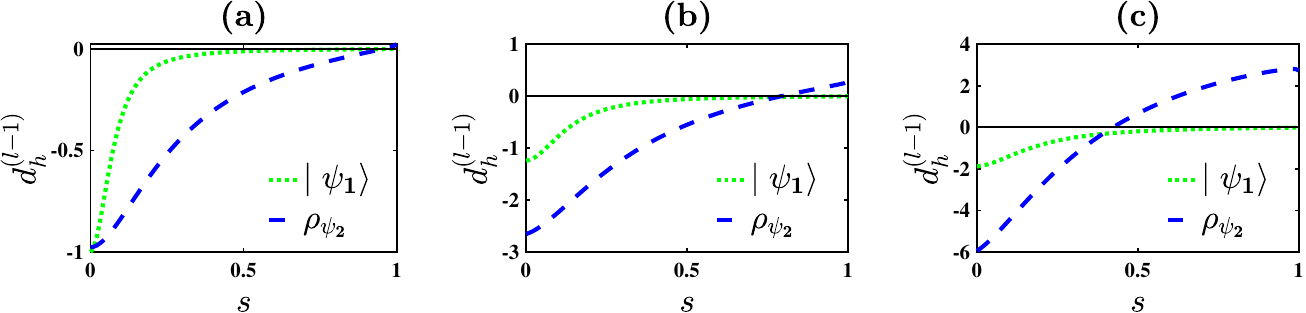}
\caption{(Color online) Comparison of $d_h^{(l-1)}$ as a function of $s$, for coherent state $\ket{\psi_1}$ and thermal state $\rho_{\psi_2}$  with (a) $|\alpha|=\bar{n}=0.1$, $l=2$, (b) $|\alpha|=\bar{n}=0.2$, $l=3$ and (c) $|\alpha|=\bar{n}=0.3$, $l=4$.}
\label{fig4}
\end{figure}

In Fig.~\ref{fig3}, the variation in antibunching of SOCS (SOTS) is shown with respect to the state variable $\alpha$ ($\bar{n}$). Here lower as well as higher-order antibunching criteria demonstrate the nonclassicality for both the states. {\color{black}It is observed that the depth of nonclassicality witness increases with the order of nonclassicality as depicted in Figures~\ref{fig3} \textcolor{black}{(a)-(c)}. This fact is consistent with the earlier observations (\cite{kishore2,kishore3}
and references therein) that higher-order nonclassicality criteria is useful in detecting weaker nonclassicality. Specifically, the depth of nonclassicality present in SOCS and SOTS, lowers as $s$ increases from 0.2 to 0.8 keeping $l$ fixed [see Figures~\ref{fig3} \textcolor{black}{(b), (e), (h)} and \ref{fig3}\textcolor{black}{(c), (f), (i)}]}.

\subsection{Higher-order sub-Poissonian photon statistics}

HOSPS is one of the important criteria of nonclassicality and sometimes gives better result than HOA. It is further investigated that HOA and HOSPS do not depend on each other \cite{kishore2} and both of them can exist irrespective of whether their lower-order counterparts exist or not \cite{nasir2}.

The generalized criteria for observing the $(l-1)$-th order sub-Poissonian photon statistics (for which $\langle(\Delta\mathcal{N})^l\rangle < \langle(\Delta\mathcal{N})^l\rangle_{\ket{\mathrm{Poissonian}}}$ is given by \cite{amit}

\begin{equation}
\mathcal{D}_h^{(l-1)} = \sum_{e=0}^{l}\sum_{f=1}^{e}S_2(e,f)^lC_e(-1)^ed_h^{(f-1)}{\langle a^\dagger a \rangle }^{l-e}\,\,<\,\,0
\label{eq14}
\end{equation}
where $S_2(e, f)=\sum_{r=0}^{f} {^fC_r}(-1)^r r^e$ is the Stirling number of second kind, $^lC_e$ is the usual binomial coefficient.
The analytic expressions of HOSPS for the SOCS and SOTS can be obtained by substituting \eqref{eq4} and \eqref{eq7} in \eqref{eq14}.

\begin{figure*}[h]
\centering
\includegraphics[scale=01]{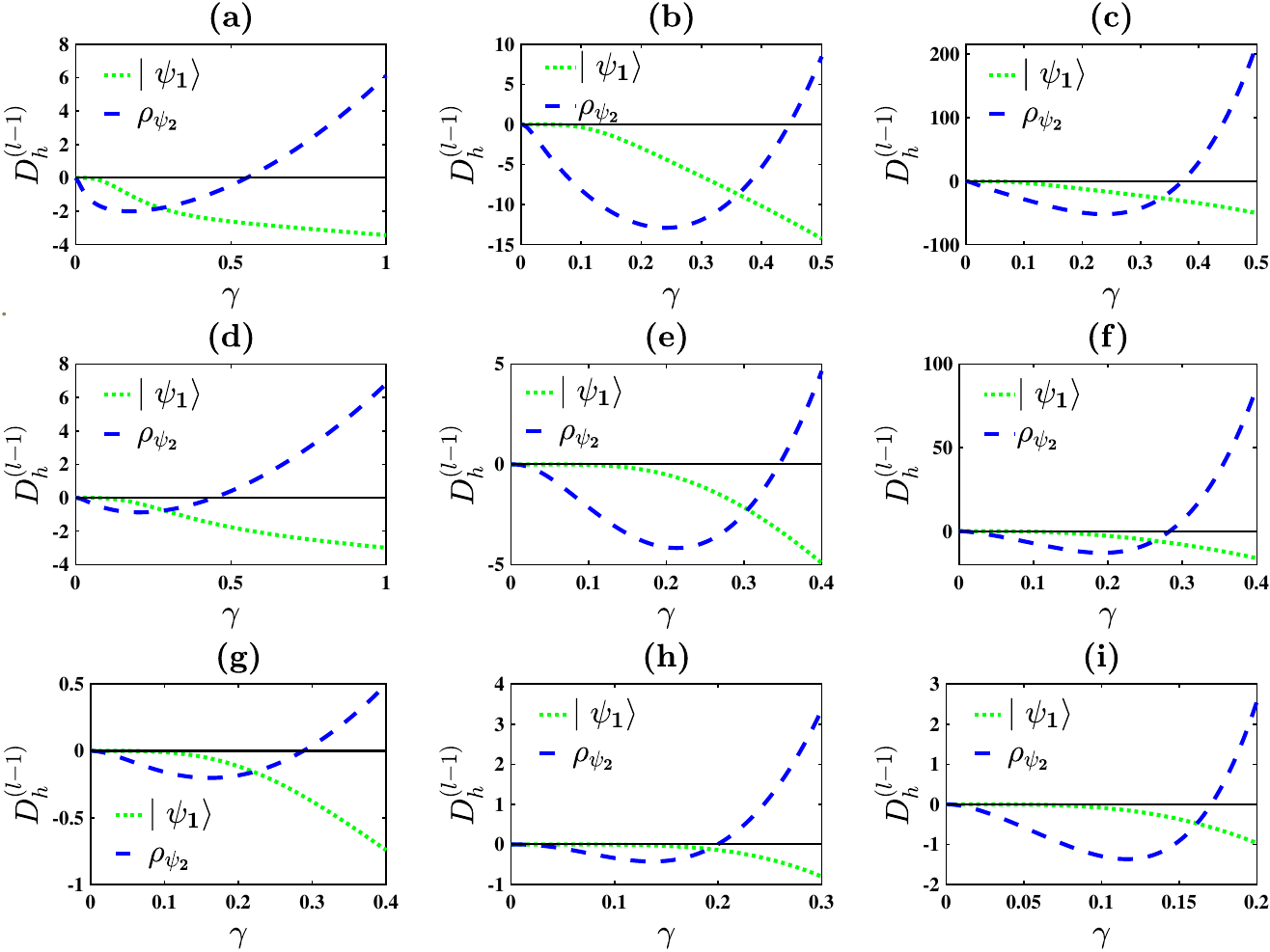}
\caption{(Color online) Comparison of $D_h^{(l-1)}$ with respect to the variable $\gamma$ which is equal to $|\alpha|$ for coherent state $\ket{\psi_1}$ and $\bar{n}$ for thermal state $\rho_{\psi_2}$ and with (a)-(c) $s=0.2$ and $l=2,3,4$, (d)-(f) $s=0.5$ and $l=2, 3, 4$, (g)-(i) $s=0.8$ and $l=2,3,4$, respectively.}
\label{fig5}
\end{figure*}
\begin{figure}[h]
\centering
\includegraphics[scale=1]{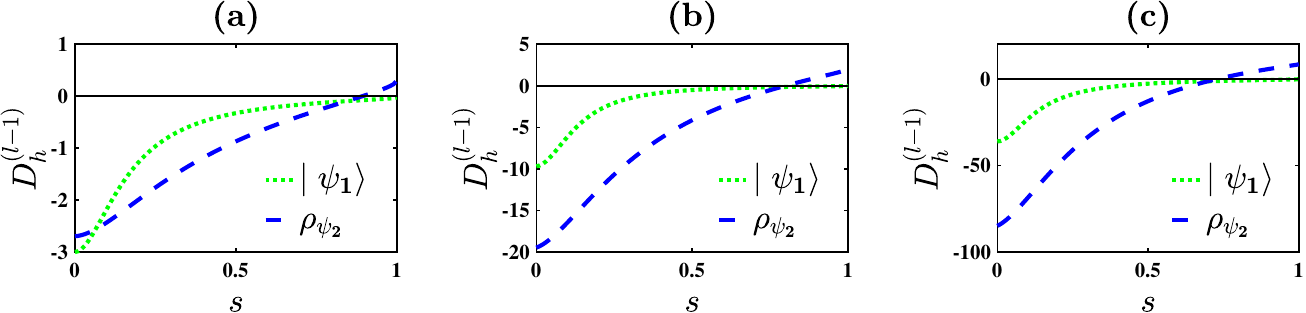}
\caption{(Color online) Comparison of $D_h^{(l-1)}$ as a function of $s$, for coherent state $\ket{\psi_1}$ and thermal state $\rho_{\psi_2}$  with $|\alpha|=\bar{n}=0.2$ and (a) $l=2$, (b) $l=3$ and (c) $l=4$.}
\label{fig6}
\end{figure}

$\mathcal{D}_h^{(l-1)}$ is plotted in Fig.~\ref{fig5} with respect to the state variable $\alpha$ ($\bar{n}$) for SUP-operated coherent (thermal) state and for different values of $l$ and $s$. HOSPS clearly detects nonclassicality for both the states under consideration. {\color{black}The lower-order sub-Poissonian photon statistics is merely present in SOTS unlike lower-order antibunching of SOTS which is detected for all values of $s$ considered here}. It is also noticed that HOSPS behaves in a similar manner as HOA does while changing $l$ and $s$.

\subsection{Higher-order squeezing}

The criterion of higher-order squeezing is given by the pioneering work of Hong and Mandel \cite{hong}. According to them, the higher-order squeezing of order $l$ ($>2$) is obtained while the $l$-th order moment of a field quadrature operator is less than the corresponding coherent state value. Hong-Mandel's criteria for higher-order squeezing can be described by the following inequality
\begin{equation}
\label{eq15}
S^{(l)} = \frac{\langle (\Delta X)^l \rangle -{\left(\frac{1}{2}\right)}_{\left(\frac{l}{2}\right)}}{{\left(\frac{1}{2}\right)}_{\left(\frac{l}{2}\right)}}\,\,<\,\,0,
\end{equation}
where $(x)_l$ is the conventional Pochhammer symbol and the quadrature variable is defined as $X = \frac{1}{\sqrt{2}}(a+a^\dagger)$. The inequality in \eqref{eq15} can also be rewritten as
\begin{equation}
\label{eq16}
\langle (\Delta X)^l \rangle\,\,<\,\,{\left(\frac{1}{2}\right)}_{\left(\frac{l}{2}\right)} = \frac{1}{2^{\frac{l}{2}}}(l-1)!!,
\end{equation}
with
\begin{eqnarray}
\label{eq17}
\langle (\Delta X)^l \rangle = \sum_{r=0}^{l}\sum_{i=0}^{\frac{r}{2}}\sum_{k=0}^{r-2i}(-1)^r\frac{1}{2^\frac{1}{2}}(2i-1)!^{2i}C_k^lC_r^rC_{2i}\langle a^\dagger +a\rangle^{l-r}\langle a^{\dagger k}a^{r-2i-k}\rangle,
\end{eqnarray}
where $l$ is an even number and
\begin{eqnarray*}
n!!=
\left\{
\begin{array}{lll}
& n(n-2)(n-4)\ldots 4.2 & \mbox{if $n$ is even},\\\\
& n(n-2)(n-4)\ldots3.1 & \mbox{if $n$ is odd},
\end{array}
\right.
\end{eqnarray*}
The analytic expression for the Hong-Mandel type HOS can be obtained by using \eqref{eq1} in \eqref{eq15}-\eqref{eq17}.
\begin{figure*}[h]
\centering
\includegraphics[scale=01]{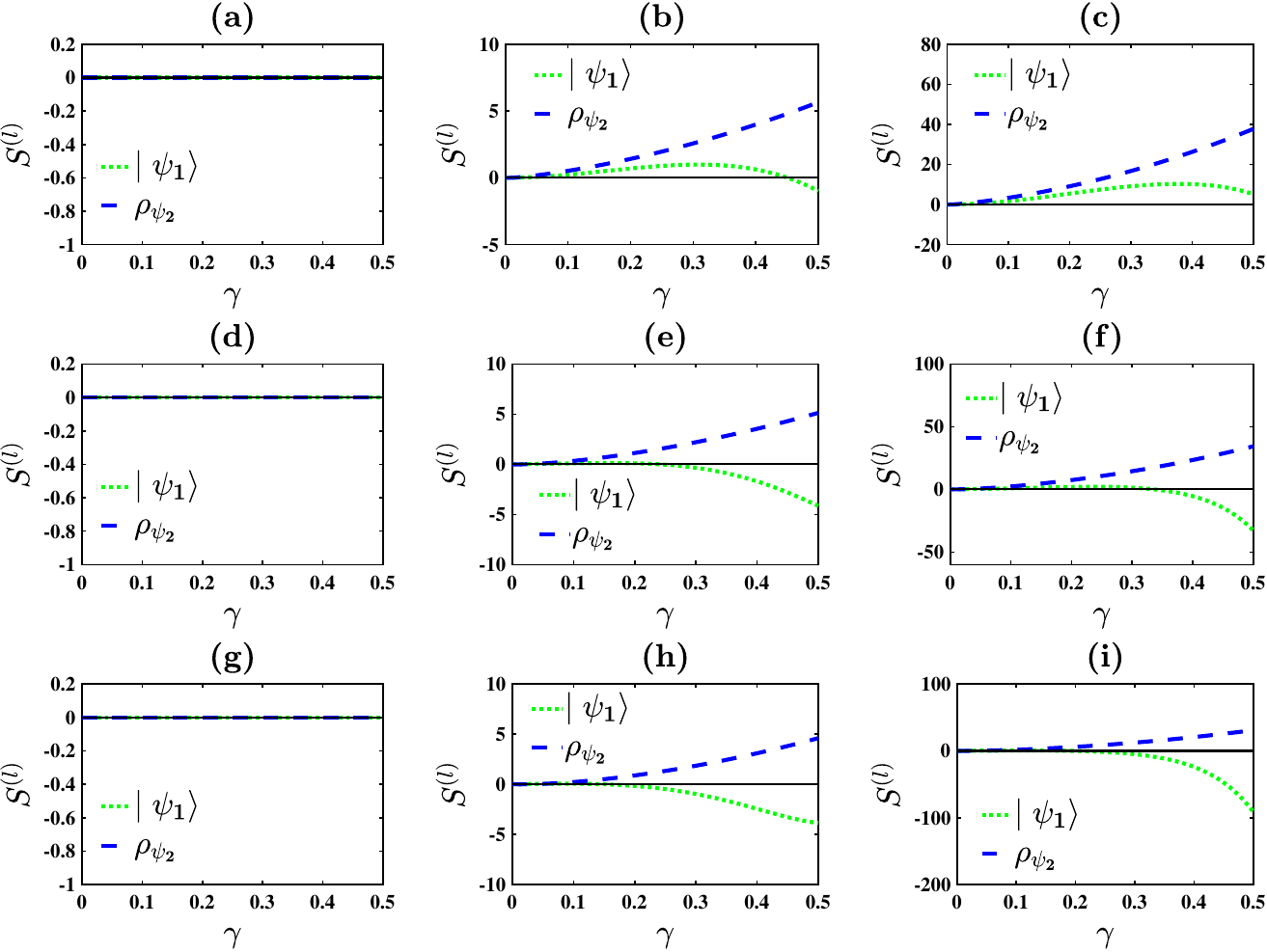}
\caption{(Color online) Comparison of $S^{(l)}$ with respect to the variable $\gamma$ which is equal to $|\alpha|$ for coherent state $\ket{\psi_1}$ and $\bar{n}$ for thermal state $\rho_{\psi_2}$ and with (a)-(c) $s=0.2$ and $l=2, 4, 6$, (d)-(f) $s=0.5$ and $l=2,4,6$, (g)-(i) $s=0.8$ and $l=2,4,6$, respectively.}
\label{fig7}
\end{figure*}
\begin{figure}[h]
\centering
\includegraphics[scale=01]{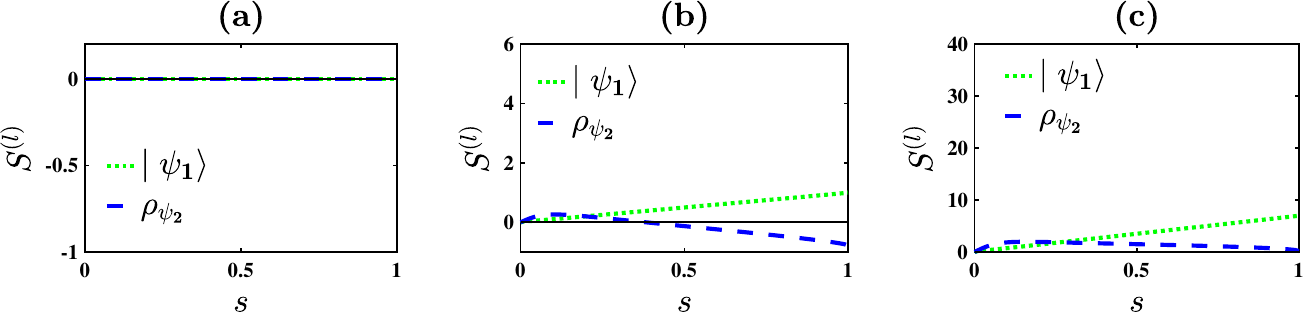}
\caption{(Color online) Comparison of $S^{(l)}$ as a function of $s$, for coherent state $\ket{\psi_1}$ and thermal state $\rho_{\psi_2}$  with $|\alpha|=\bar{n}=0.01$ and (a)  $l=2$, (b) $l=4$ and (c) $l=6$, respectively.}
\label{fig8}
\end{figure}

Fig.~\ref{fig7} illustrates the existence of Hong-Mandel type HOS in the superposed states, assuming arbitrary values of the state variables $\alpha$ or $\bar{n}$ and for different orders of squeezing ($l=2,\,4,\,6$). Here the lower-order squeezing criteria fails to detect the nonclassicality of SUP-operated states [see Figures~\ref{fig7} \textcolor{black}{(a), (d), (g)}] while its higher-order counterpart hardly witnesses the nonclassicality of SOTS. Thus HOS is not much advantageous for identifying the nonclassicalities of the considered states.

\subsection{$Q$ function}

A direct phase space description of a quantum mechanical system is not possible due to the uncertainty principle. This fact leads to the construction of quasiprobability distributions which are very useful in quantum mechanics as they provide a quantum classical correspondence and facilitate the calculation of quantum mechanical averages in close analogy to classical phase space averages \cite{kishore3}. One such quasiprobability distribution is $Q$ function, and zeros of this function are a signature of nonclassicality \cite{husimi}. The Husimi-$Q$ function is defined as
\begin{equation}\nonumber
Q = \frac{1}{\pi}\bra{\beta}\rho\ket{\beta},
\end{equation}
where $\ket{\beta}$ is the usual coherent state. A simple calculation finds the Husimi-$Q$ for SOCS and SOTS as following:
\begin{eqnarray}
\begin{array}{rcl}
\label{eq20}
Q_1 & = & \frac{1}{\pi}N_1^{-1} |\langle{\beta|\psi_1}\rangle|^2 \\\\
& = & \frac{1}{\pi}N_1^{-1}\left|s+(s+t)\alpha\beta^*\right|^2 e^{-|\alpha|^2-|\beta|^2+\alpha\beta^*+\alpha^*\beta},
\end{array}
\end{eqnarray}
\begin{eqnarray}
\begin{array}{rcl}
\label{eq19}
Q_2 &=& \frac{1}{\pi} N_2^{-1}\langle{\beta|\rho_{\psi_2}|\beta}\rangle\\\\
&=&\frac{1}{\pi(1+\bar{n})}N_2^{-1}e^{-\frac{|\beta|^2}{1+\bar{n}}}\left[\left\{s+(s+t)\left(\frac{\bar{n}|\beta|^2}{1+\bar{n}}\right)
\right\}^2+(s+t)^2\left(\frac{\bar{n}|\beta|^2}{1+\bar{n}}\right)\right]
\end{array}
\end{eqnarray}

%

The zeros of Husmi $Q$ function for the state $\ket{\psi_1}$ can be calculated by using the condition $s+(s+t)\alpha\beta^* = 0$ which gives  $s=\frac{\alpha\beta^*}{\sqrt{1+2\alpha\beta^* + 2(\alpha\beta^*)^2}}$.


\begin{figure}[h]
\centering
\includegraphics[scale=01]{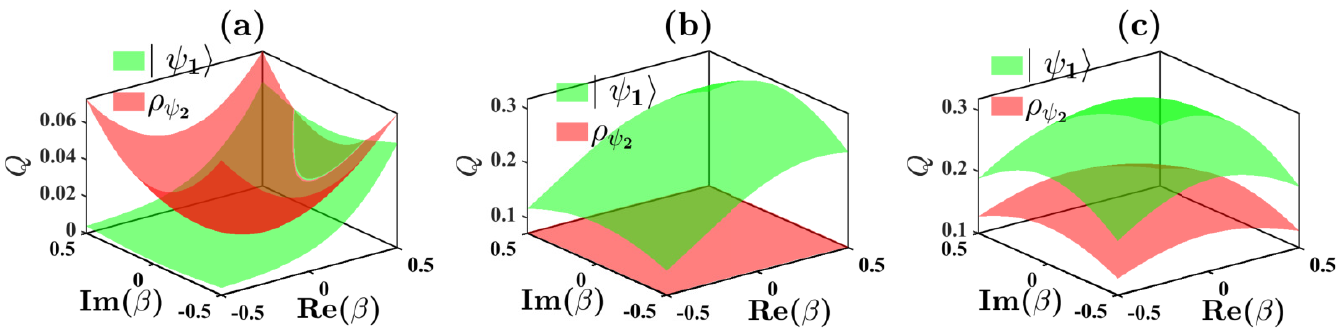}

\caption{(Color online) Comparison of Husimi-$Q$ as a function of $\beta$, for different values of state variables $|\alpha|$ for coherent state $\ket{\psi_1}$ and $\bar{n}$ for thermal state $\rho_{\psi_2}$ with (a) $|\alpha|=\bar{n}=1$ and $s=0.2$, (b) $|\alpha|=\bar{n}=0.1$ and $s=0.5$, (c) $|\alpha|=\bar{n}=0.01$ and $s=0.8$.}

\label{fig9}
\end{figure}

{\color{black}Fig.~\ref{fig9} shows that, in general we have failed to observe the nonclassical features reflected beyond moments based nonclassicality criteria through a
quasiprobability distribution, i.e., zeros of the $Q$ function. Particularly when $s=0.2$ and state variable $\alpha=1$ ($\bar{n}=1$), the $Q$ function has zero for SOCS (SOTS)}. Thus Husimi-$Q$ depicts the nonclassical nature of both the states in some specific region.

\subsection{Agarwal-Tara criterion}

Agarwal and Tara proposed a moment based criteria for witnessing the nonclassical character of a given quantum state \cite{girish}. They defined $A_3$ which consists of the moments of the number distribution $\mu_j$ and the normal ordered moments $m_j$. The analytic expression of $A_3$ in terms of these higher ordered moments is \cite{priya}
\begin{equation}
\label{eq21}
A_3 = \frac{{\mbox{det}\,\,m}^{(3)}}{{\mbox{det}\,\,\mu}^{(3)} - {\mbox{det}\,\,m}^{(3)}}\,\, <\,\, 0,
\end{equation}
where
\begin{eqnarray*}
m^{(3)} =
\begin{bmatrix} 1 & m_1 & m_2\\
m_1 & m_2 & m_3\\m_2 & m_3 & m_4
\end{bmatrix},
\end{eqnarray*}
and
\begin{eqnarray*}
\mu^{(3)} =
\begin{bmatrix} 1 & \mu_1 & \mu_2\\
\mu_1 & \mu_2 & \mu_3\\\mu_2 & \mu_3 & \mu_4
\end{bmatrix}.
\end{eqnarray*}
The matrix elements are defined by $m_j = \langle{a^{\dagger j}a^j}\rangle$ and  $\mu_j = \langle{(a^\dagger a)^j}\rangle$.

\begin{figure}[h]
\centering
\includegraphics[scale=01]{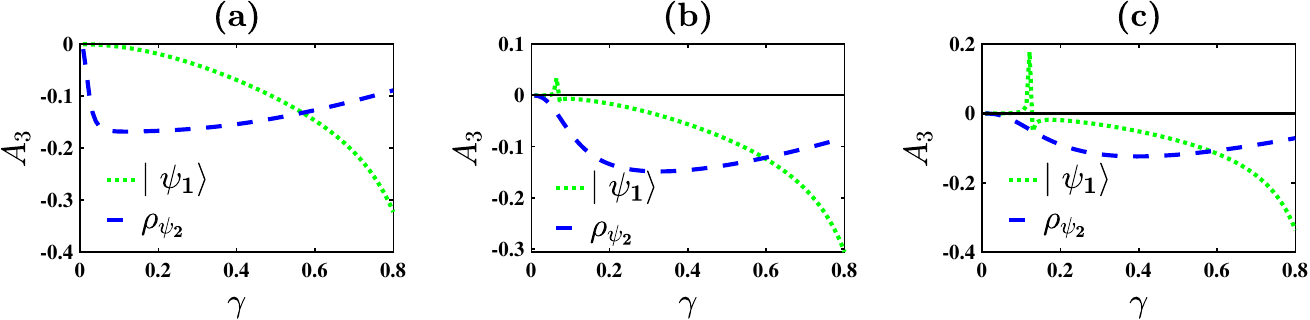}
\caption{(Color online) Variation of Agarwal-Tara parameter with respect to the state parameter $\gamma$ which is equal to  $|\alpha|$ for coherent state $\ket{\psi_1}$ and $\bar{n}$ for thermal state $\rho_{\psi_2}$ with (a) $s=0.01$ (b) $s=0.1$ (c) $s=0.2$, respectively.}
\label{fig10}
\end{figure}
\begin{figure}[h]
\centering
\includegraphics[scale=1]{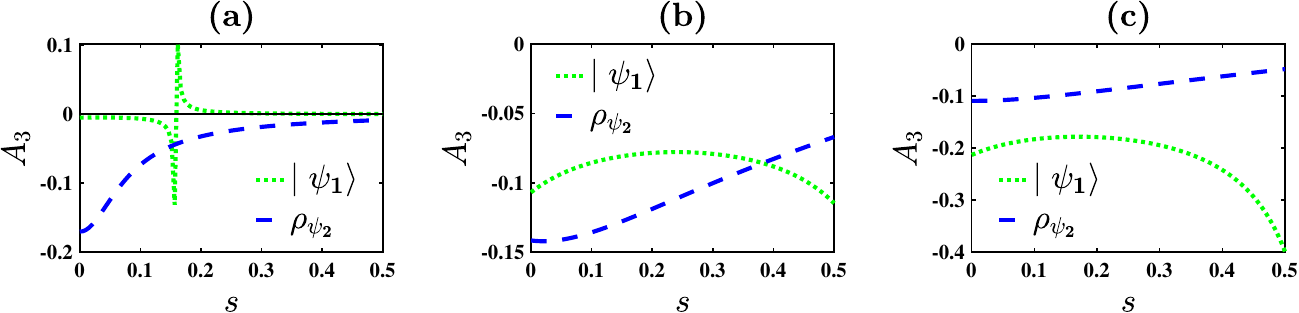}
\caption{(Color online) Variation of Agarwal-Tara parameter with respect to $s$ for coherent state $\ket{\psi_1}$ and thermal state $\rho_{\psi_2}$ where (a) $|\alpha|=\bar{n}=0.1$ (b) $|\alpha|=\bar{n}=0.5$ and (c) $|\alpha|=\bar{n}=0.7$, respectively.}
\label{fig11}
\end{figure}

For a classical-like coherent state, the value of $A_3$ is 0 while for the most nonclassical Fock state, it is -1. Thus for any nonclassical state, $A_3$ is negative and bounded  below by -1 (if $A_3=-1$, the corresponding state is called maximally nonclassical). To investigate the nonclassicality of the considered state via the Agarwal-Tara parameter, we have plotted $A_3$ [see Fig.~\ref{fig10}] as a function of $\alpha$ ($\bar{n}$) for SOCS (SOTS). {\color{black}Here $A_3$ lies between 0 and -1 for all values of $s$. The zero value of $A_3$ of SOCS for the displacement parameter $\alpha=0$ is consistent with the fact that vacuum state is a classical state having non-negative $P$-function. For some specific mean values, Agarwal-Tara criterion for SOCS shows a singularity [cf. Figures~\ref{fig10}-\ref{fig11}], which indicates that the probability of photons coming bunched is equal to that of coming independently. Nonclassicality reflected through $A_3$ parameter decreases (increases) while increasing the SUP-operator parameter $s$, and the mean photon number $\alpha$ ($\bar{n}$) for SOCS (SOTS).}

\subsection{Klyshko's criterion}

One of the witness to check the nonclassicality of any quantum state was given by Klyshko \cite{klyshko}, which is based only on three consecutive photon-number probabilities. If $p_m = \langle{{m}|\rho|{m}}\rangle$ is the photon-number probability of a state with density matrix $\rho$, then the Klyshko's inequality can be written as
\begin{equation}
\label{eq22}
B(m) = (m+2)p_mp_{m+2} - (m+1)(p_{m+1})^2\,\, <\,\,0
\end{equation}
Using \eqref{eq1} and \eqref{eq22}, the functions $B(m)$ for both the states are obtained as follows:
\begin{eqnarray}
\begin{array}{lcl}
\label{eq23}
B(m)_1 & = & N_1^{-2}|\alpha|^{4m+4}\left[(m+2)\left\{s+(s+t)m\right\}^2\left\{s+(s+t)(m+2)\right\}^2\right.\\
& & \left.-(m+1)\left\{s+(s+t)(m+1)\right\}^4\right]
\end{array}
\end{eqnarray}
\begin{eqnarray}
\begin{array}{lcl}
\label{eq24}
B(m)_2 & = & N_2^{-2}\left(\frac{1}{1+\bar{n}}\right)^2 \left(\frac{\bar{n}}{1+\bar{n}}\right)^{2m+2}\left[(m+2)\left\{s+(s+t)m\right\}^2\right.\\
& & \left.\times\left\{s+(s+t)(m+2)\right\}^2-(m+1)\left\{s+(s+t)(m+1)\right\}^4\right]
\end{array}
\end{eqnarray}
where $B(m)_1$, $B(m)_2$ are corresponding to SOCS and SOTS respectively.
\begin{figure}[h]
\centering
\includegraphics[scale=1]{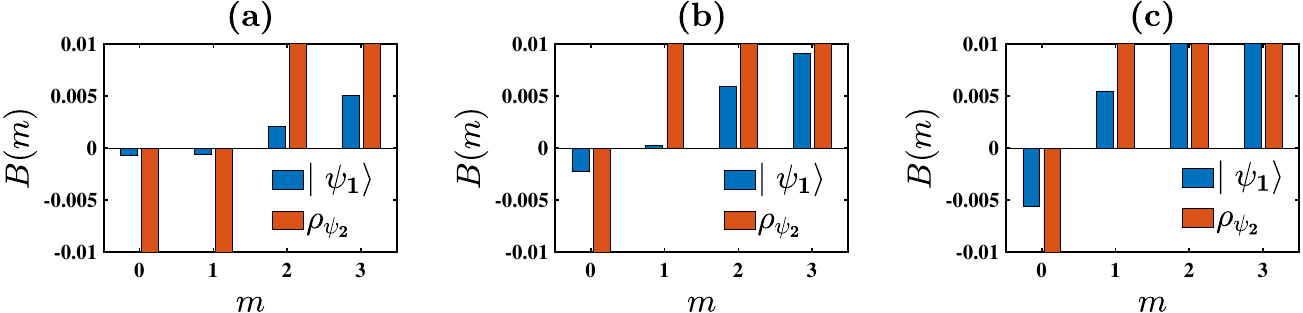}
\caption{(Color online) Illustration of Klyshko's criterion $B(m)$ as a function of $m$ for coherent state $\ket{\psi_1}$ and thermal state $\rho_{\psi_2}$ with (a) $|\alpha|=\bar{n}=2$, $s=0.2$, (b) $|\alpha|=\bar{n}=1.5$, $s=0.5$, (c) $|\alpha|=\bar{n}=1$, $s=0.8$.}
\label{fig12}
\end{figure}

In this criterion, very less information is needed with respect to the other conditions as probability of only three successive photon numbers is sufficient to investigate this
nonclassical property. The negative values of $B(m)$ indicate the nonclassicality of the corresponding state. It is clear from Fig.~\ref{fig12} that Klyshko's condition signifies nonclassicality only for $m=0,\,1$ and as $s$ increases from 0.2 to 0.8, $B(m)$ fails to detect the nonclassicality.

\begin{figure}[h]
\centering
\includegraphics[scale=0.65]{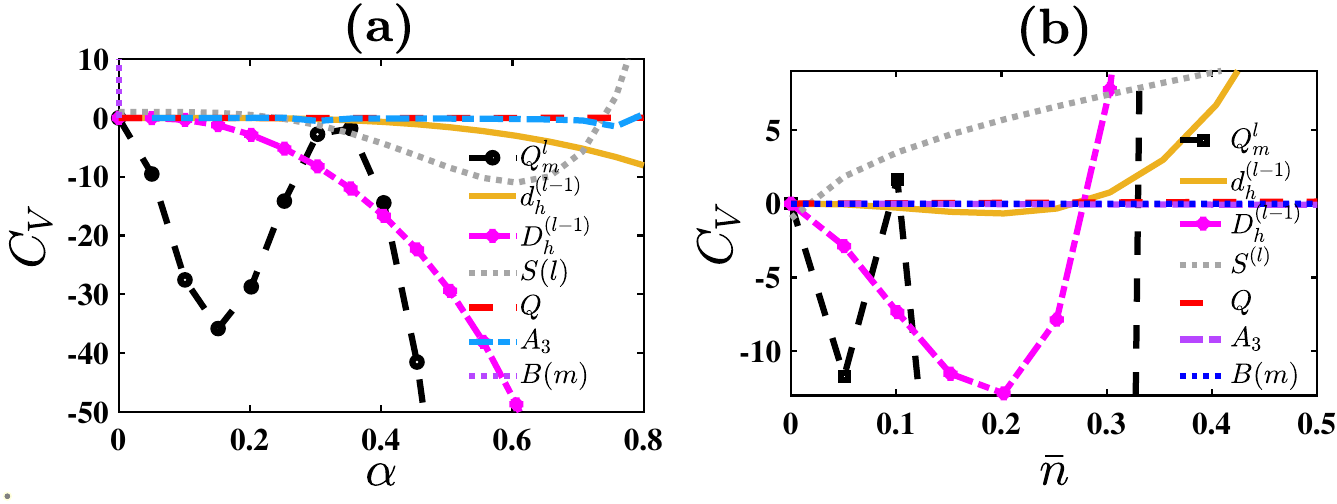}
{\color{black}
\caption{(Color online) Illustration of all criteria values ($C_v$) as a function of (a) $|\alpha|$ for coherent state $\ket{\psi_1}$ and (b) $\bar{n}$ for thermal state $\rho_{\psi_2}$ with arbitrary parametric values $s=0.5$, $l=4$ and for Klyshko's criterion $m=4$.}
\label{fig13}}
\end{figure}

{\color{black}Fig.~\ref{fig13} describes a comparative plot for all the seven criteria studied
so far, as a function of $\alpha$ for coherent state $\ket{\psi_1}$ and $\bar{n}$ for thermal state $\rho_{\psi_2}$ with arbitrary parametric values $s=0.5$ and $l=4$. Almost all of the witnesses signify the nonclassical nature of SUP-operated coherent state and SUP-operated thermal state except $S^{(l)}$ for SOTS.}

{\color{black}\section{Effect of non-unit quantum efficiency}

Light comes in the form of photons. The quantum efficiency states that for every photon coming in there is a probability that the photon will generate
an electron which is measured by the detector. An ideal single-photon detector is a measurement device that perfectly discriminates the photon-number
states, i.e. a box that produces $n$ clicks when the number state $\ket{n}$ comes in \cite{datta}. The vacuum state $\ket{0}$ can be absolutely extricated from all the other Fock states $\ket{n_{n\geq 1}}$ by a unit-efficiency and noise-free detector \cite{zhang}. To account for the non-unit efficiency $\eta$, the mode
$a$ is sent through a beam-splitter with transmission probability $\eta$. The output modes are labelled as $a$ and
$l$ and the corresponding operator is given by $C_\eta =
e^{{\sqrt{1-\eta}/\eta}a l^{\dagger}} {\eta}^{\frac{1}{2} a^{\dagger}a}\ket{0_l}$. Thus an inefficient detector is well-described
by the operator \cite{gisin}
\begin{eqnarray}
D(\eta) = (1-\eta)^{a^\dagger a}
\end{eqnarray}
Applying the operator $D(\eta)$ on SOCS and SOTS, we obtain $D(\eta)$-operated SOCS (DSOCS) and $D(\eta)$-operated SOTS (DSOTS) as follows
\begin{eqnarray}
\left.
\begin{array}{lcl}
\label{eqi1}
\ket{\psi_{1}^\eta} & = & (N_1^\eta)^{-1/2}D(\eta)A\ket{\alpha},\\\\
\rho_{\psi_{2}^\eta} & = & (N_2^\eta)^{-1}D^\dagger(\eta)A^\dagger\rho_n A D(\eta),
\end{array}
\right\}
\end{eqnarray}
where
\begin{eqnarray*}
N_1^\eta & = & \left[\left\{s+(s+t)|\alpha|^2\right\}^2+(s+t)^2|\alpha|^2\right]+\left[\left\{s+(s+t)(1-\eta)|\alpha|^2\right\}^2+(s+t)^2(1-\eta)|\alpha|^2\right]e^{-\eta|\alpha|^2}\\
& & +\left[\left\{s+(s+t)(1-\eta)^2|\alpha|^2\right\}^2+(s+t)^2(1-\eta)^2|\alpha|^2\right]e^{(\eta^2-2\eta)|\alpha|^2},
\end{eqnarray*}
and
\begin{eqnarray*}
N_2^\eta & = & \frac{1}{(1+\bar{n})}\left[\left[\left\{s+(s+t)\mu^2\right\}^2+(s+t)^2 \mu^2\right]e^\mu-2\left[\left\{s+(s+t)\theta^2\right\}^2+(s+t)^2 \theta^2\right]e^\theta\right.\\
& & \left.+\left[\left\{s+(s+t)\zeta^2\right\}^2+(s+t)^2\zeta^2\right]e^\zeta\right],
\end{eqnarray*}
are the normalization constants for DSOCS and DSOTS, respectively and $\mu=\frac{\bar{n}}{1+\bar{n}}$, $\theta=\mu(1-\eta)$, $\zeta=\mu(1-\eta)^2$. The expectation of $a^{\dagger m}a^n$ with respect to DSOCS is given by
\begin{eqnarray}\nonumber
\label{eqi2}
{\langle{a^{\dagger m}a^n}\rangle}_1^\eta & = & \bra{\psi_1^\eta} a^{\dagger m}a^n \ket{\psi_1^\eta}\\\nonumber
& = & (N_1^\eta)^{-1}\alpha^{* m}\alpha^n\left[\left\{\big(s+(s+t)|\alpha|^2\big)^2+(s+t)^2|\alpha|^2\right\}+\left\{(1-\eta)^m+(1-\eta)^n\right\}\right.\\\nonumber
& & \left.\times\left\{\big(s+(s+t)
(1-\eta)|\alpha|^2\big)^2+(s+t)^2(1-\eta)|\alpha|^2\right\}e^{-\eta|\alpha|^2}+\big(1-\eta\big)^{m+n}\right.\\
& & \left.\times\left\{\big(s+(s+t)(1-\eta)^2|\alpha|^2\big)^2+(s+t)^2(1-\eta)^2|\alpha|^2\right\}e^{(\eta^2-2\eta)|\alpha|^2}\right],
\end{eqnarray}
and with respect to DSOTS, it is
\begin{eqnarray}\nonumber
\label{eqi3}
{\langle{a^{\dagger m}a^n}\rangle}_2^\eta & = & (N_2^\eta)^{-1}\,\mathrm{tr}(a^{\dagger m}a^n D^\dagger(\eta)A^\dagger\rho_r AD(\eta))\\
& = & \delta_{mn} (N_2^\eta)^{-1}\sum_r \frac{a_r~r!}{(r-n)!}(s+(s+t)r)^2\left\{1-(1-\eta)^r\right\}^2,
\end{eqnarray}
where $\delta_{m n}$ is the usual delta function. Using \eqref{eqi2} and \eqref{eqi3}, the analytic expressions of several higher-order nonclassicality criterion for DSOCS and DSOTS are evaluated and plotted as follows:}

%

\begin{figure*}[h]
\label{figi1}
\centering
\includegraphics[scale=1]{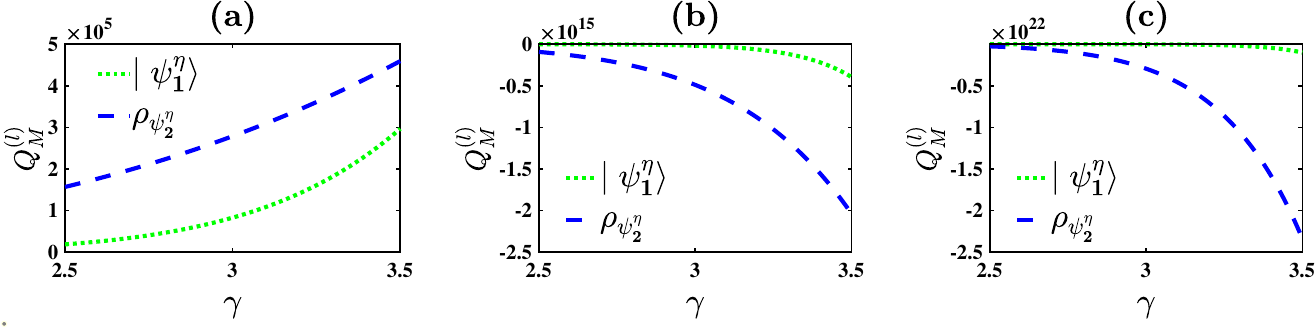}

{\color{black} \caption{(Color online) Comparison of $Q_M^{(l)}$ with respect to the variable $\gamma$ which is equal to $|\alpha|$ for coherent state $\ket{\psi_{1}^\eta}$ and $\bar{n}$ for thermal state $\rho_{\psi_{2}^\eta}$ and with $s=0.2$ (a) $l=2$, $\eta=0.25$, (b) $l=5$, $\eta=0.5$, (c) $l=7$, $\eta=0.75$, respectively.}}

\end{figure*}

\begin{figure*}[h]
\centering
\includegraphics[scale=1]{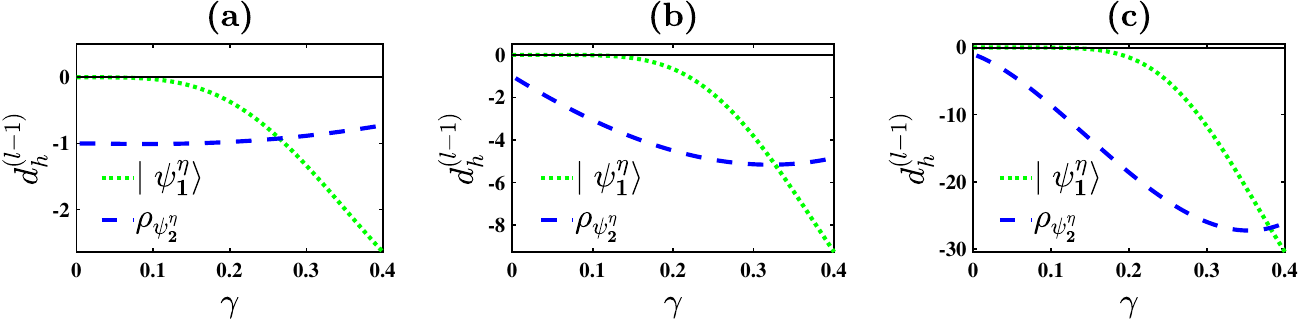}

{\color{black} \caption{(Color online) Comparison of $d_h^{(l-1)}$ with respect to the variable $\gamma$ which is equal to $|\alpha|$ for coherent state $\ket{\psi_{1}^\eta}$ and $\bar{n}$ for thermal state $\rho_{\psi_{2}^\eta}$ and with $s=0.2$ (a) $l=2$, $\eta=0.75$, (b) $l=3$, $\eta=0.5$,(c) $l=4$, $\eta=0.25$, respectively.}}
\label{figi2}
\end{figure*}

\begin{figure*}[h]
\centering
\includegraphics[scale=1]{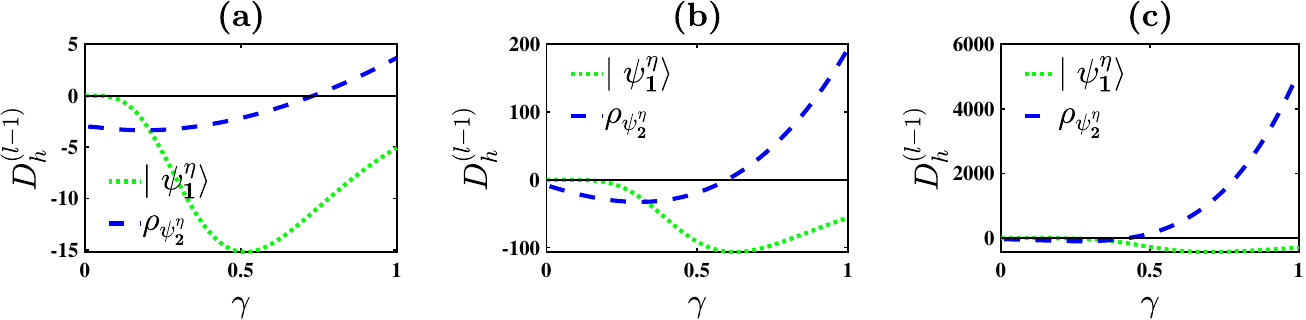}

{\color{black} \caption{(Color online) Comparison of $D_h^{(l-1)}$ with respect to the variable $\gamma$ which is equal to $|\alpha|$ for coherent state $\ket{\psi_{1}^\eta}$ and $\bar{n}$ for thermal state $\rho_{\psi_{2}^\eta}$ and with $s=0.2$ and (a) $l=2$, $\eta=0.3$, (b) $l=3$, $\eta=0.6$ and (c) $l=4$, $\eta=0.9$, respectively.}}
\label{figi3}
\end{figure*}

\begin{figure*}[h]
\centering
\includegraphics[scale=01]{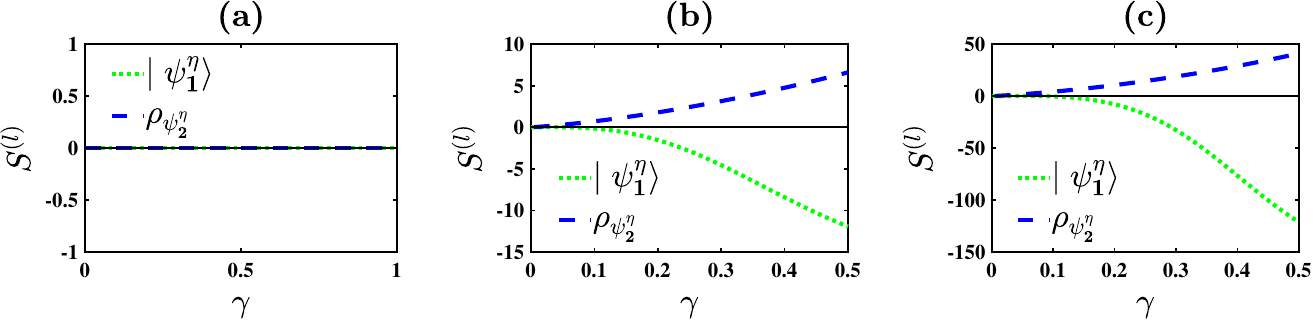}

{\color{black}
\caption{(Color online) Comparison of $S^{(l)}$ with respect to the variable $\gamma$ which is equal to $|\alpha|$ for coherent state $\ket{\psi_{1}^\eta}$ and $\bar{n}$ for thermal state $\rho_{\psi_{2}^\eta}$ and with $s=0.2$ and (a) $l=2$, $\eta=0.4$, (b) $l=4$, $\eta=0.6$ and (c) $l=6$, $\eta=0.8$, respectively.}}
\label{figi7}
\end{figure*}

\begin{figure}[h]
\centering
\includegraphics[scale=01]{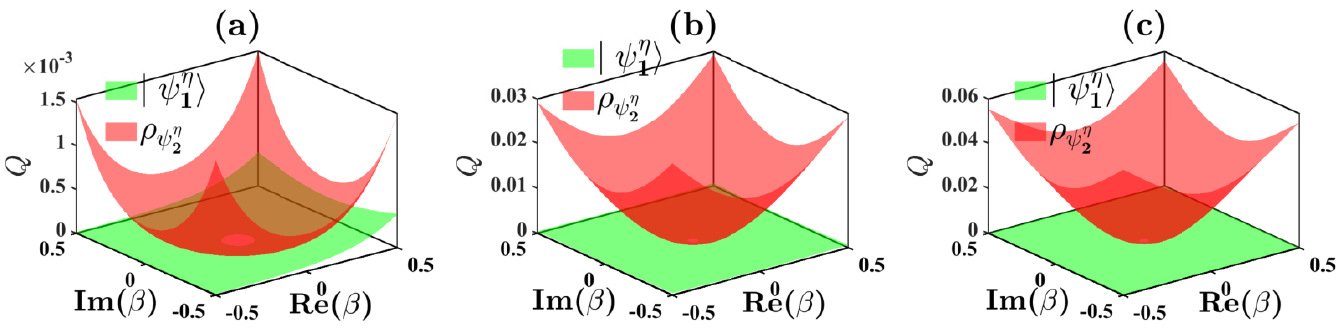}

{\color{black}
\caption{(Color online) Comparison of Husimi-$Q$ as a function of $\beta$, for different values of state variables $|\alpha|$ for coherent state $\ket{\psi_{1}^\eta}$ and $\bar{n}$ for thermal state $\rho_{\psi_{2}^\eta}$ with (a) $|\alpha|=\bar{n}=1$, $s=0.2$ and $\eta=0.1$, (b) $|\alpha|=\bar{n}=0.1$, $s=0.5$ and $\eta=0.4$, (c) $|\alpha|=\bar{n}=0.01$, $s=0.8$ and $\eta=0.7$, respectively.}}

\label{figi5}
\end{figure}

\begin{figure}[h]
\centering
\includegraphics[scale=01]{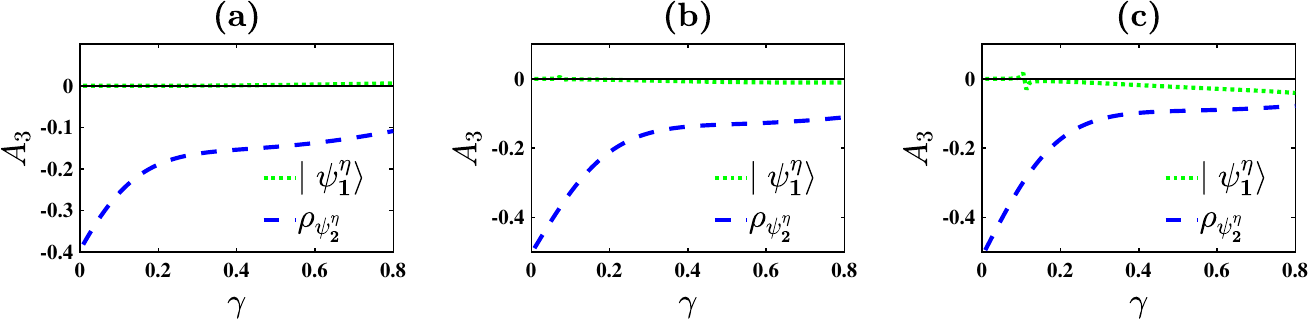}

{\color{black}
\caption{(Color online) Variation of Agarwal-Tara parameter with respect to the state parameter $\gamma$ which is equal to $|\alpha|$ for coherent state $\ket{\psi_{1}^\eta}$ and $\bar{n}$ for thermal state $\rho_{\psi_{2}^\eta}$ with (a) $s=0.01$, $\eta=0.8$, (b) $s=0.1$, $\eta=0.5$, (c) $s=0.2$, $\eta=0.2$, respectively.}}
\label{figi6}
\end{figure}

\begin{figure}[h]
\centering
\includegraphics[scale=01]{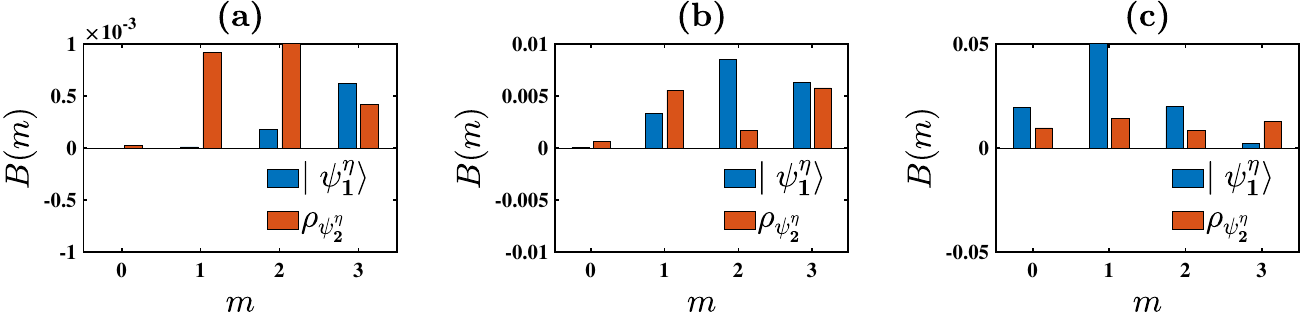}
\label{fign7}
{\color{black}
\caption{(Color online) Illustration of Klyshko's criterion $B(m)$ as a function of $m$ for coherent state $\ket{\psi_{1}^\eta}$ and thermal state $\rho_{\psi_{2}^\eta}$ with (a) $|\alpha|=\bar{n}=2$, $s=0.2$, $\eta=0.4$, (b) $|\alpha|=\bar{n}=1.5$, $s=0.5$, $\eta=0.6$, (c) $|\alpha|=\bar{n}=1$, $s=0.8$, $\eta=0.8$, respectively.}}

\end{figure}

{\color{black}Here, one can visualize that due to the imperfections of real detector, the depth of nonclassicality of the superposed states increases in case of  higher-order photon statistics. The same is true for HOA, HOSPS and Husimi-$Q$. We may note that HOS is not observed for the SUP-operated thermal state while non-unit quantum efficiency is taken into account. Additionally, the negative peaks in the values of $B(m)$ disappear. That means, the typical realistic detector can control higher-order nonclassicalities of the states studied here.
}

\section{Conclusion}
\label{sec4}

In the present work, we have introduced a quantum state by applying a combination of product of two operators $a$ and $a^{\dagger}$ to a coherent (thermal) state $\ket{\psi_1}$ ($\rho_{\psi_2}$) with $s^2+t^2=1$. The scalars $s$ and $t$ act as control parameters for manipulation of the nonclassical character of the output state. We have focused on the higher-order nonclassical features of the states. A set of different measurement techniques is used to check the existence of higher-order nonclassicality in the superposed states. It is found that higher-order measurements (except HOS and Klyshko) perform better to detect the nonclassicality of SOCS and SOTS.  {\color{black}In brief, various types of higher-order nonclassical features of SUP-operated coherent as well as thermal states are investigated and their variations with experimentally controllable parameters are studied. We have shown that the depth of higher-order nonclassicality witnesses increase with the order of nonclassicality which establishes the fact that higher-order nonclassicality criteria has an advantage in detecting weaker
nonclassicality. The effect of non-unit quantum efficiency of photon detector is also discussed which ensures that the proposed work can be realized experimentally in the near future. We can wind up this article with a hope that the allowed control over the depth of nonclassicality witnesses will be put to further applications in the context of quantum information processing.}

\begin{center}
\textbf{ACKNOWLEDGEMENT}
\end{center}
Deepak's work is supported by the Council of Scientific and Industrial Research (CSIR), Govt. of India (Award no. 09/1256(0006)/2019-EMR-1).


\end{document}